\documentclass{PoS}
\usepackage{amssymb}

\title{First HAWC Spectra of Galactic Gamma-ray Sources Above 100 TeV and the Implications for Cosmic-ray Acceleration}

\ShortTitle{HAWC Spectra above 100 TeV}

\author{\speaker{Kelly Malone} for the HAWC Collaboration\footnote{for collaboration list see PoS(ICRC2019)1177 or https://www.hawc-observatory.org/collaboration/icrc2019.php} \\
        Los Alamos National Laboratory\\
        E-mail: \email{kmalone@lanl.gov}}
        
\abstract{We present the first catalogs of the highest-energy (above 56 TeV and 100 TeV) gamma-ray sources seen by the High Altitude Water Cherenkov (HAWC) Observatory. The wide field-of-view of HAWC naturally lends itself to unbiased all-sky surveys and newly developed event-by-event gamma-ray energy reconstruction algorithms have allowed unprecedented energy resolution. The sources presented here are the highest-energy sources ever detected. All are coincident with known lower-energy gamma-ray sources within our Galaxy. These objects may have implications for the sources of Galactic cosmic rays; since Galactic CRs have been observed up to PeV energies, sources accelerating particles to these energies must exist. These sources, called ``PeVatrons'', would have corresponding hard gamma-ray spectra that extend to high energies without any spectral break or cutoff. We will present measurements of the spectra of these highest-energy gamma-ray sources and discuss if any of them can be identified as PeVatron candidates.}

\FullConference{36th International Cosmic Ray Conference -ICRC2019-\\
		July 24th - August 1st, 2019\\
		Madison, WI, U.S.A.}

\begin{document}

\section{Introduction}
Although the spectrum of cosmic rays (CRs) has been accurately measured over several decades of energy, the actual sources of cosmic rays are still unknown.  The CR spectrum is Galactic-dominated up to at least the ``knee'' feature ($\sim$1 PeV)\cite{Gaisser2016}, so one would expect that at least a few sources within the Galaxy are able to accelerate CRs up to or past this energy.  These hypothetical sources are called ``PeVatrons''. To date, only one such potential source has been found (in the Galactic Center, by the HESS Collaboration\cite{Abramowski2016}). While this source may have been more active in the past, it currently does not have a high enough rate of particle acceleration to substantially contribute to the population of Galactic cosmic rays.

Part of the difficulty in identifying cosmic ray sources is due to their charged nature: CRs bend in magnetic fields on their way to the Earth and therefore do not point back to their sources. It is advantageous to instead study gamma rays. CRs interact with the ISM or nearby photon fields and create neutral pions. These neural pions will subsequently decay to gamma rays. Since gamma rays are neutral, they do not bend in magnetic fields and are an excellent probe of cosmic ray sources. The resultant gamma rays are approximately an order of magnitude less energetic than the parent CR.  For a 1 PeV CR, the gamma ray will be $\sim$100 TeV\cite{Aharonian2013}. The gamma-ray spectrum of a PeVatron is expected to be fairly hard ($\sim$E$^{-2.0}$) and extend to at least tens of TeV without any exponential cutoff or spectral break. 

Observations at the highest gamma-ray energies ($>$ 50 TeV) have been scarce.  Fluxes at these energies are extremely low.  The High Altitude Water Cherenkov (HAWC) Observatory is an extensive air shower array experiment located at 4100m in the state of Puebla, Mexico. It has an extremely high duty cycle ($> 95\%$) and wide instantaneous field-of-view ($\sim$2 steradian)\cite{Smith2015,Abeysekara2017a}. This makes it an ideal instrument to search the sky for the highest-energy emission.

HAWC is capable of observing sources between -26 degrees and +64 degrees in declination. It is optimized for the energy range above 300 GeV, with detections above 100 TeV possible.  Above 10 TeV, it is the most sensitive currently-operating gamma-ray observatory in the world. 

Note that merely observing $>$ 100 TeV gamma rays does not conclusively identify a PeVatron. Gamma rays can also be created through leptonic channels, such as Inverse Compton scattering. Due to the Klein-Nishina effect, this leptonic component becomes suppressed starting arounds tens of TeV, which results in a changing spectral index with energy\cite{Moderski2005}.  Several leptonic PWN have been detected at the highest energies - the Crab Nebula extends above 100 TeV\cite{Abeysekara2019,Amenomori2019}, and HESS J1825-134 extends to at least 70 TeV\cite{Abdalla2019}.

This proceeding describes the construction of catalogs of gamma-ray sources above 56 TeV and 100 TeV using data from the HAWC observatory. 

\section{Gamma-ray energy estimation}
The HAWC Collaboration uses two recently developed energy estimation algorithms that vastly increase the dynamic range of the experiment. Previously, the percentage of the photomultiplier tubes (PMTs) hit during an event was used as a proxy for energy (see \cite{Abeysekara2017a} for details). This led to a loss of dynamic range above $\sim$10-20 TeV, as every PMT in the array is typically hit for showers above this energy. 

This proceeding uses the new energy estimation algorithm known as the ``ground parameter'' (GP). This method uses the charge measured 40 meters from the air shower axis along with the zenith angle of the event to estimate energy. It is discussed in-depth in \cite{Abeysekara2019}. The GP has a low energy bias above $\sim$30 TeV, making it a good choice to search for PeVatrons and other objects that emit at the highest energies.
\section{Verification of the energy estimator on the Crab Nebula}
The performance of the GP energy estimator has been extensively studied on the Crab Nebula using 837.2 days of HAWC data\cite{Abeysekara2019}. The spectrum is well-fit to a log parabola that continues emitting past 100 TeV:

\begin{equation}
\frac{dN}{dE} = \phi_0 \left(\frac{E}{E_0} \right)^{-\alpha-\beta \mathrm{ln}(E/E_0)},
\end{equation}
where $\phi_0$ = (2.35$\pm$0.04$^{+0.20}_{-0.21}) \times$ 10$^{-13}$ (TeV cm$^2$ s)$^{-1}$, $E_0$ is the pivot energy and is fixed at 7 TeV, $\alpha$ = 2.79$\pm$0.02$^{+0.01}_{-0.03}$ and $\beta$=0.10$\pm$0.01$^{+0.01}_{-0.03}$. For each parameter, the first set of uncertainties are statistical and the second set are systematic.

Above 100 TeV in reconstructed energy, the statistical significance of the Crab Nebula is $\sim$4$\sigma$, where significance is defined as the square root of the test statistic that is obtained from a likelihood ratio test. This is the among the highest-energy detections of any gamma-ray source to date. See \cite{Linnemann2019} for a discussion of the highest-energy part of the Crab Nebula spectrum. 

HAWC's previously published Crab spectrum\cite{Abeysekara2017a} was only valid up to 40 TeV. The lower-energy part of this new spectrum agrees with the previously published one. The maximum detected energy is increased by approximately a factor of three. 

\begin{figure}
\centering
\includegraphics[width=0.8\textwidth]{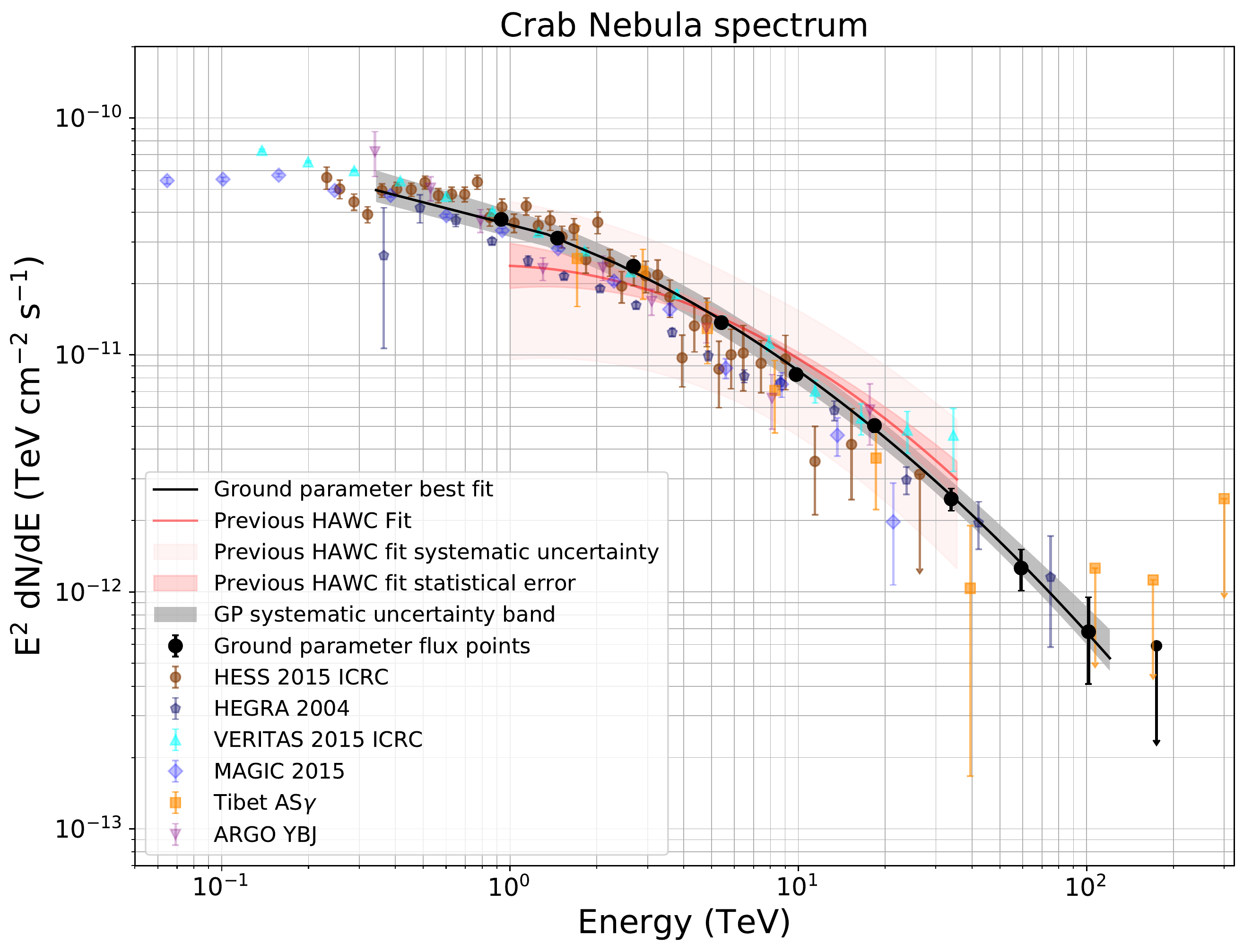}
\caption{The black flux points show the spectrum of the Crab Nebula obtained with the ground parameter method described above. The error bars are statistical uncertainties only. The solid black line is the forward-folded log-parabola best fit, and the shaded gray band is the systematic uncertainty on that fit. For comparison, the HAWC fit from \cite{Abeysekara2017a} is shown, as are results from selected other experiments. The references for the other experiments are: HESS \cite{Holler2015}, VERTIAS\cite{Meagher2015}, MAGIC\cite{Aleksic2015}, Tibet AS$\gamma$\cite{Amenomori2015}, ARGO YBJ\cite{Bartoli2015}, HEGRA\cite{Aharaonian2004} }
\label{fig:crab}
\end{figure}

In the regime where HAWC's energy range overlaps with that of imaging atmospheric Cherenkov telescopes (a few TeV), the measured fluxes agree well (within $\sim$20$\%$).  The Crab Nebula spectrum, including a comparison to relevant IACTs, can be seen in Figure \ref{fig:crab}.

\section{High-energy catalog construction}
Here, a catalog of the highest-energy sources is constructed using 1038.8 days of data collected between June 2015 and July 2018. This search is conducted above two energy thresholds: 56 TeV and 100 TeV. The binning scheme and gamma/hadron separation are described in \cite{Abeysekara2019}.

Significance maps for each chosen energy threshold are created using a likelihood framework\cite{Vianello2015}. A hypothetical point source with a power-law E$^{-2}$ spectrum (as would be expected for a PeVatron) is moved across the sky and the test statistic is computed as compared to the background-only hypothesis. This process is then repeated twice more to cover cases where the hypothetical source is spatially extended (disks with radii of 0.5 degree and 1.0 degree are assumed).  There are a total of 6 maps: the three different source morphologies at each of the two different energy thresholds.

The catalog is then built by searching all maps for local maxima where the test statistic is $>$ 25. The same technique used in constructing HAWC's second all-energy catalog (\cite{Abeysekara2017b}, hereafter known as the ``2HWC catalog") is followed here: sources must be separated from each other by a valley of $\Delta\sqrt{TS} > 2$.  

An exceptionally bright source may appear in the catalog search up to six times (the three different morphologies multiplied by the two energy thresholds).  The location of the hotspot may move slightly between these maps. To obtain the one definitive source location and extension, we float the right ascension, declination, and extension of each source in the $>$ 56 TeV map under the assumption of a ``PeVatron-like'' spectrum ($E^{-2.0}$). These results are reported here.

After identifying areas exhibiting high-energy emission, the integral flux and spectrum is computed using the same likelihood framework. The source location is fixed at the fitted $>$ 56 TeV location.
\section{Results} 
Table \ref{table:results} gives the location and extension of each high-energy source. There are nine sources with significant emission above 56 TeV in reconstructed energy. Eight of these sources are within one degree of the Galactic plane; the ninth is the Crab Nebula. All of the sources within one degree of the Galactic plane are extended in apparent size. 

All sources are coincident with known lower-energy sources. Notably, eight of the nine sources are within 0.5$^\circ$ of sources from the 2HWC\cite{Abeysekara2017b} catalog.   The ninth source, eHWC J1839-057, is nearly a degree from the nearest 2HWC source, 2HWC J1837-065. There was emission seen at the location of eHAWC J1839-057 in the 2HWC catalog, but it was not flagged as its own source due to the requirement that sources be separated by a valley of $\Delta\sqrt{TS}>2$. By applying a higher-energy threshold as is done, we can see that the region likely contains two sources with different energy thresholds that were not originally identified as such due to the limitations of the catalog construction method.  

\begin{table}
\centering
\title{Sources seen above 56 TeV}
\begin{tabular}{|c||c|c|c|c|c|c|}
\hline
Source name & RA ($^{\circ}$) & Dec ($^{\circ}$) & Ext. ($^{\circ}$) & Nearest & 2HWC& $>$ 100\\
 &  & &  & 2HWC & dist. ($^{\circ}$) & TeV \\
\hline\hline
eHWC J0534+220 & 83.61 $\pm$ 0.02 & 22.00 $\pm$ 0.03 & PS & J0534+220 & 0.02 &  \\
eHWC J1809-193 & 272.46  $\pm$ 0.13 & -19.34  $\pm$ 0.14 & 0.34  $\pm$ 0.13 &  J1809-190 & 0.30 & \\
eHWC J1825-134 & 276.40  $\pm$ 0.06 & -13.37  $\pm$ 0.06 & 0.36  $\pm$ 0.05 &  J1825-134 & 0.07 & \checkmark \\
eHWC J1839-057 & 279.77  $\pm$ 0.12 & -5.71  $\pm$ 0.10 & 0.34  $\pm$ 0.08 &  J1837-065 & 0.96 & \\
eHWC J1842-035 & 280.72  $\pm$ 0.15 & -3.51  $\pm$ 0.11 & 0.39  $\pm$ 0.09 & J1844-032 & 0.44 &   \\
eHWC J1850+001 & 282.59  $\pm$ 0.21 & 0.14  $\pm$ 0.12 & 0.37  $\pm$ 0.16 &  J1849+001 & 0.20 & \\
eHWC J1907+063 & 286.91  $\pm$ 0.10 & 6.32  $\pm$ 0.09 & 0.52  $\pm$ 0.09 &  J1908+063 & 0.16 & \checkmark  \\
eHWC J2019+368 & 304.95  $\pm$ 0.07 & 36.78  $\pm$ 0.04 & 0.20  $\pm$ 0.05 &  J2019+367 & 0.02 & \checkmark \\
eHWC J2030+412 & 307.74  $\pm$ 0.09 & 41.23  $\pm$ 0.07 & 0.18  $\pm$ 0.06 & J2031+415 & 0.34 & \\
\hline
\end{tabular}
\caption{The sources seen above 56 TeV. Sources are found in a blind catalog search similar to the 2HWC catalog construction method. Uncertainties are statistical only. A Gaussian morphology is assumed for all sources except for the Crab Nebula; the column labeled ``Ext" is the Gaussian width. The ``PS'' entry for the Crab denotes that it is best modeled as a point source. The sources that have TS $>$ 25 emission above 100 TeV in reconstructed energy have checkmarks in the last column. }
\label{table:results}
\end{table}

\begin{figure}
\centering
\includegraphics[width=\textwidth]{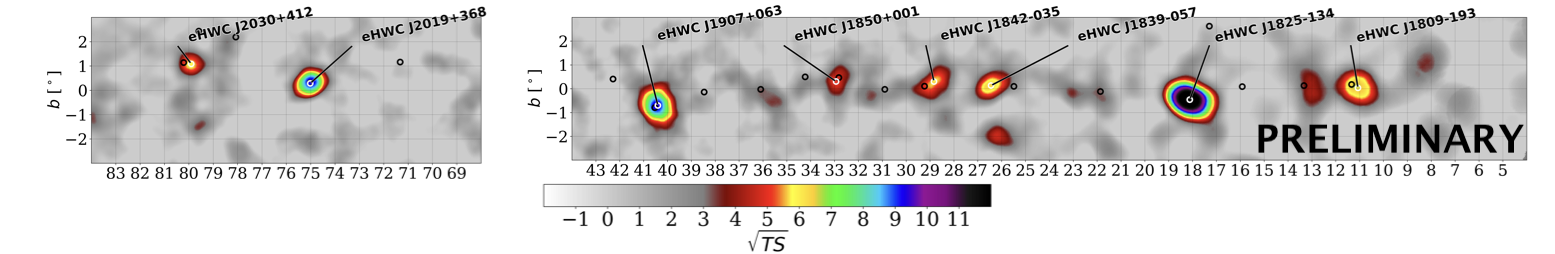}
\caption{The Galactic plane above 56 TeV in reconstructed energy. Since all of these sources are extended in apparent size, the morphology in this figure is assumed to be a disk with a radius of 0.5 degree.  White open circles are $>$ 56 TeV hotspots (defined as emission with TS $>$ 25).  Black circles denote sources from the 2HWC catalog.} 
\label{fig:gr56}
\end{figure}

\begin{figure}
\centering
\includegraphics[width=\textwidth]{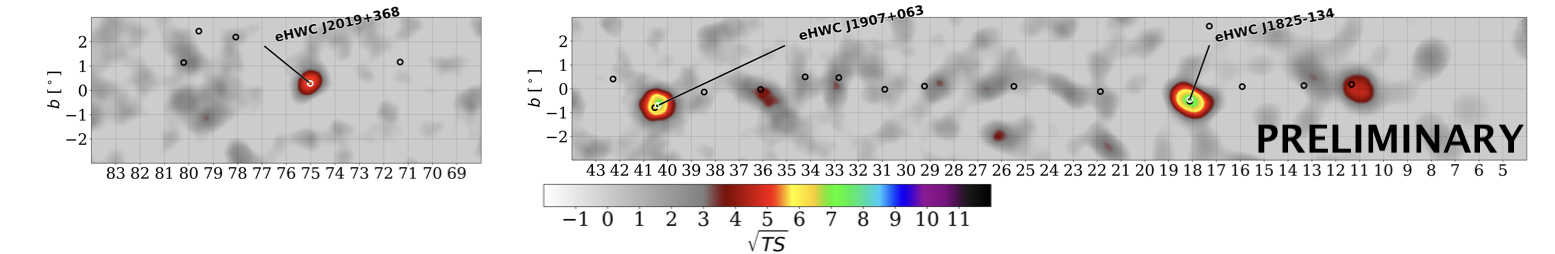}
\caption{The same as Figure \ref{fig:gr56}, but for an energy threshold of 100 TeV.} 
\label{fig:gr100}
\end{figure}

\begin{figure}
\includegraphics[width=0.5\textwidth]{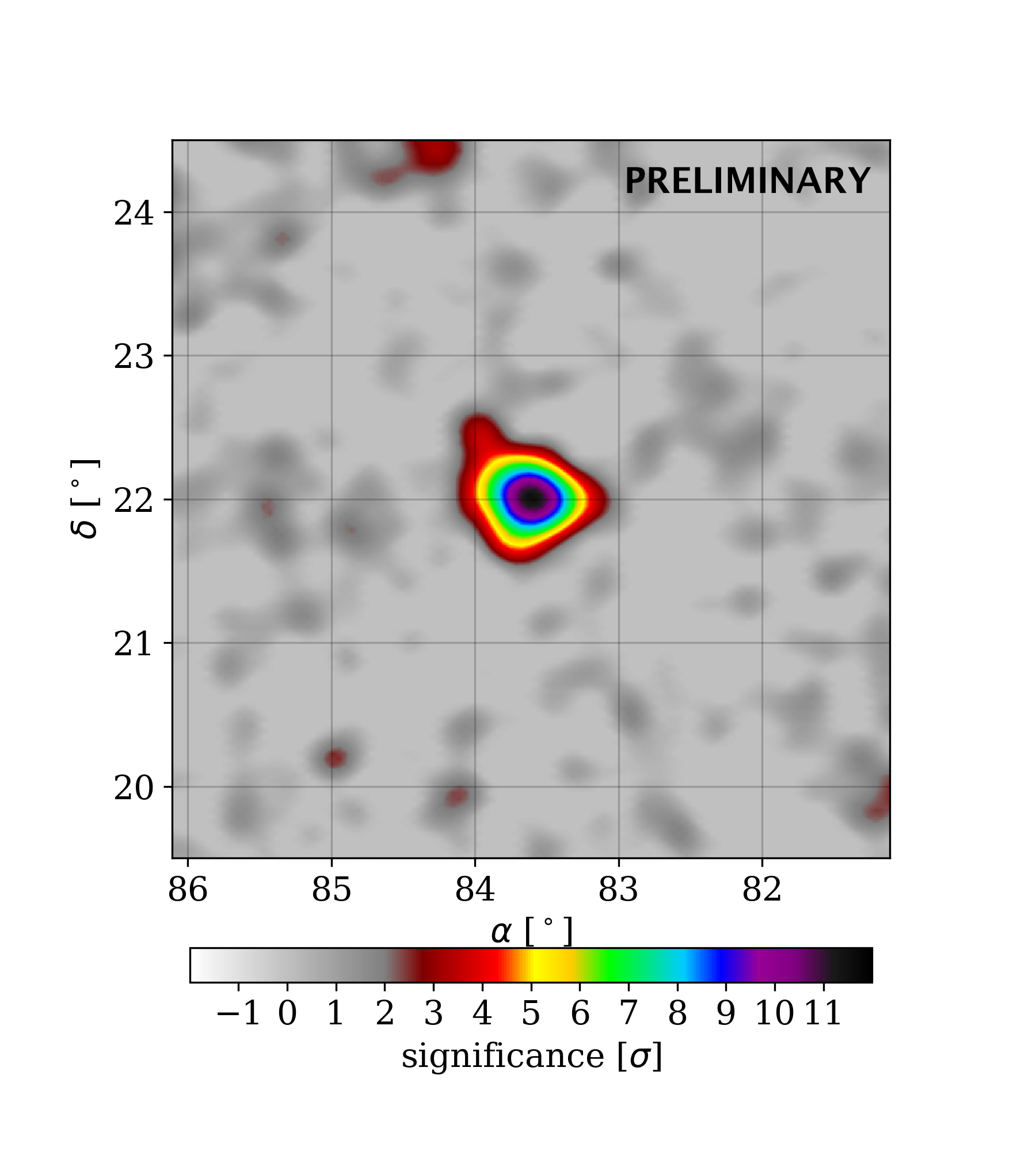}
\includegraphics[width=0.5\textwidth]{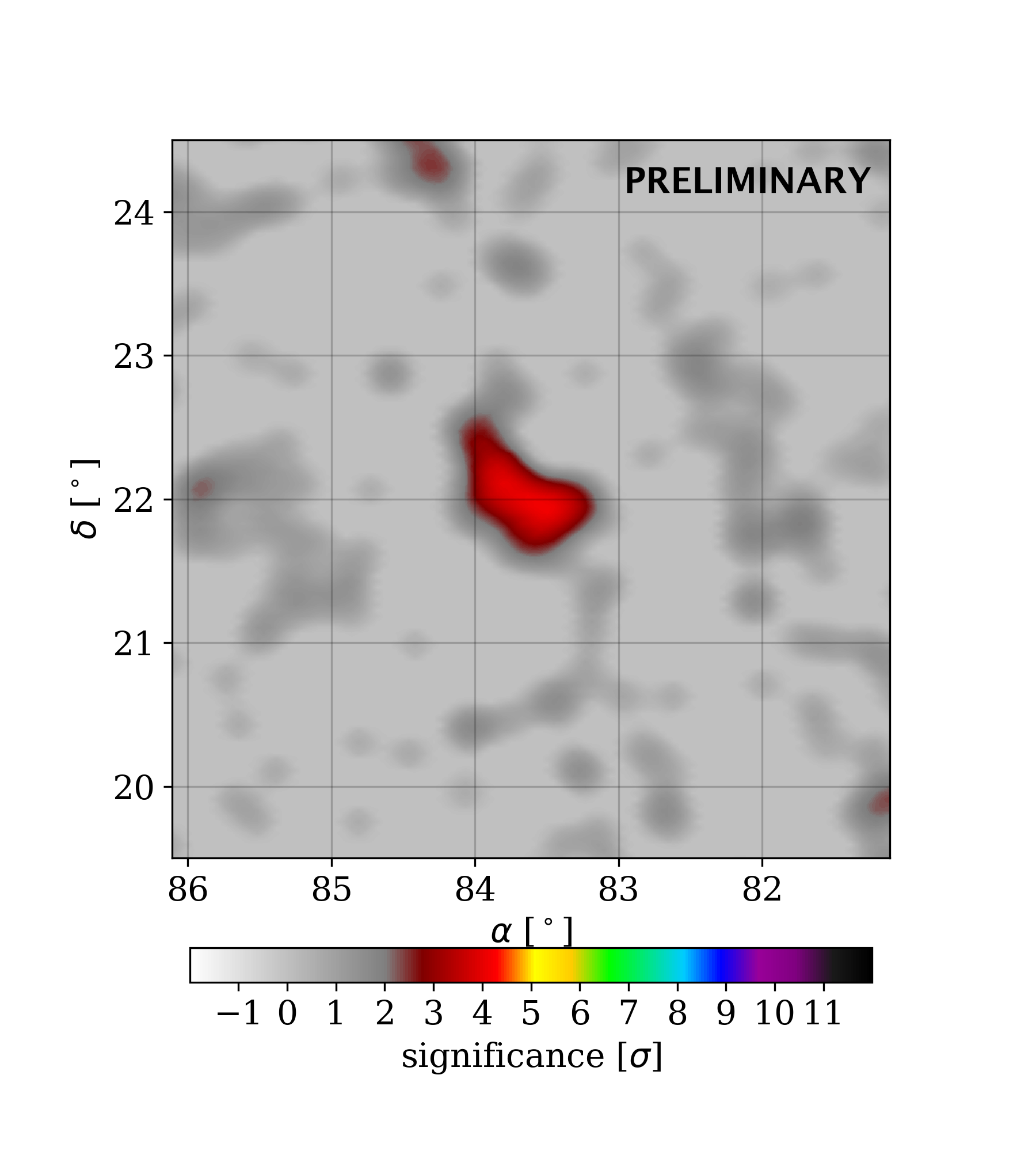}
\caption{Significance map of the Crab Nebula above 56 TeV in reconstructed energy (left) and above 100 TeV in reconstructed energy (right). A point source morphology is assumed.  The maximum significance is 11.59$\sigma$ above 56 TeV in reconstructed energy and 4.18$\sigma$ above 100 TeV in reconstructed energy. }
\label{fig:crabHE}
\end{figure}

Three of these nine sources continue to show significant emission (TS $>$ 25) above 100 TeV in reconstructed energy.  These are among the highest-energy sources ever detected, by any messenger.   Figures \ref{fig:gr56} and \ref{fig:gr100} show significance maps of the inner Galactic plane above 56 TeV and 100 TeV in reconstructed energy, respectively.   Figure \ref{fig:crabHE} contains significance maps of the Crab Nebula above these two energy thresholds.

Most of these sources are located within one degree of the Galactic plane. This region is extremely crowded with gamma-ray sources and there is the potential for contamination of the source flux from nearby sources. Multi-source and multi-component fits are not considered here; see \cite{Hona2019,ChadICRC,PacoICRC} for more sophisticated analyses of some of these regions.  Likewise, potential contributions from Galactic diffuse emission and/or unresolved sources are also not included in this analysis. 

Due to this limitation, it is impossible to compute spectra for some of these sources with this analysis technique. For example, the eHWC J2030+412 region overlaps both a PWN and the TeV counterpart of Fermi cocoon\cite{Ackermann2011,Hona2019}. Without a multi-source fit, the spectrum of the cocoon region is contaminated by the PWN. Other sources, such as eHWC J1839-057, are located close to sources that only emit at lower energies, which has the potential to contaminate the lower-energy part of the spectrum. For these sources, we do not calculate the spectra over the entire energy range but instead only quote an integral flux above 56 TeV. Since there is less source contamination above this energy, this number is less prone to uncertainties. 

The three sources that emit above 100 TeV are located in more isolated regions where contamination is not as big of a concern. We do compute spectra for these three sources.  All are well-fit to either a power-law with an exponential cutoff or a log parabola (Equations \ref{eq:plc} and \ref{eq:lp}). 

\begin{equation}
\frac{dN}{dE} = \phi_0 \left( \frac{E}{10 \mathrm{ TeV}} \right)^{-\alpha} e^{-E/E_{cut}}
\label{eq:plc}
\end{equation}

\begin{equation}
\frac{dN}{dE} = \phi_0 \left( \frac{E}{\mathrm{10 TeV}} \right)^{-\alpha - \beta \mathrm{ln}(E/E_0)}
\label{eq:lp}
\end{equation}

With the exception of the Crab Nebula, a Gaussian morphology is assumed when fitting the spectrum. The Crab Nebula is seen as a point source. The best-fit parameters for these spectra  will be given in a forthcoming publication.
\section{Testing the effect of high-energy bin migration} 
There is not a perfect 1:1 correspondence between the estimated energy given by the ground parameter algorithm and the energy from the Monte Carlo simulation. The bias and energy resolution of the energy estimator must be taken into account - some events in a given energy bin may be either higher- or lower-energy events that have been mis-reconstructed and migrated into the bin of interest.  Since astrophysical sources emit with roughly power-law spectra with negative spectral indicies, there are naturally more lower-energy events. If even a small number of those events are mis-reconstructed, they may migrate into a higher-energy bin and dwarf the number of true higher-energy events.

For the three sources where spectra are computed, the effect of this bin migration has been quantified by convolving a step function at 56 TeV with the best-fit spectral model and re-fitting. In all cases, the spectral fit from either Equation \ref{eq:plc} or \ref{eq:lp} is preferred over the version with the hard cutoff by at least 6$\sigma$. This process is repeated with a hard cutoff at 100 TeV; the nominal spectral fit is preferred over the hard cutoff at 100 TeV by at least 2.6$\sigma$ for all sources.
\section{Discussion}
In general, it is hard to conclusively determine the emission mechanisms contributing to these sources. Almost all of the sources discussed above are close to ATNF radio pulsars\cite{Manchester2005} and may be TeV halos or PWN. This would point to a dominantly leptonic origin of the emission. However, many of these pulsars are also fairly young (20-30 kyr). Therefore, there may be substantial contributions from a hadronic supernova remnant\cite{Linden2017}. 

The introduction to this proceeding noted that PeVatrons are expected to extend to $\sim$100 TeV without any spectral cutoff or break. Although all of the sources have either an exponential cutoff or a curved spectrum, this does not immediately disqualify them from being PeVatrons. Pair production on the interstellar radiation field and the CMB could lead to a steepening in the observed spectra, mimicking a cutoff in the emission spectra\cite{Porter2018}. 

More detailed studies are needed before the emission mechanisms can be conclusively determined. Since all of these sources are extended, one interesting study would be to assume diffusion from the center of the source and investigate whether particles could conceivably diffuse across the source within the electron cooling time.  

Multi-messenger and multi-wavelength studies will be important in disentangling emission mechanisms.  Most notably, observation of neutrinos coincident with any of these sources would unambiguously identify the source as having a hadronic component. Although a sub-dominant fraction of the astrophysical neutrinos observed by IceCube are expected to be Galactic in origin\cite{Aartsen2017}, a joint analysis with the data set presented here and the IceCube dataset could potentially be very interesting. 


\bibliographystyle{ICRC}
\bibliography{bib}

\end{document}